# Ubiquitous strong electron phonon coupling at the interface of FeSe/SrTiO$_3$

Chaofan Zhang[1,2], Zhongkai Liu[3], Zhuoyu Chen[1,2], Yanwu Xie[1,2], Ruihua He[4], Shujie Tang[1,2], Junfeng He[1,2], Wei Li[1,2], Tao Jia[1,2], Slavko. N. Rebec[1,2], Eric Yue Ma[1,2], Hao Yan[1,2], Makoto Hashimoto[5], Donghui Lu[5], Sung-Kwan Mo[6], Yasuyuki Hikita[1], Robert G. Moore[1,2], Harold Y. Hwang[1,2], Dunghai Lee[7] & Zhixun Shen[1,2,*]

[1]*Stanford Institute for Materials and Energy Sciences, SLAC National Accelerator Laboratory, 2575 Sand Hill Road, Menlo Park, California 94025, USA*

[2]*Geballe Laboratory for Advanced Materials, Departments of Physics and Applied Physics, Stanford University, Stanford, California 94305, USA*

[3]*School of Physical Science and Technology, ShanghaiTech University, Shanghai 200031, P.R.China*

[4]*Department of Physics, Boston College, Chestnut Hill, Massachusetts 02467, USA*

[5]*Stanford Synchrotron Radiation Lightsource, SLAC National Accelerator Laboratory, 2575 Sand Hill Road, Menlo Park, California 94025, USA*

[6]*Advanced Light Source, Lawrence Berkeley National Laboratory, Berkeley, California 94720, USA*

[7]*Department of Physics, University of California at Berkeley, Berkeley, California 94720, USA*

*\* To whom correspondence should be addressed: zxshen@stanford.edu*

**The high temperature superconductivity in single-unit-cell (1UC) FeSe on SrTiO$_3$ (STO)(001) and the observation of replica bands by angle-resolved photoemission spectroscopy (ARPES) have led to the conjecture that the coupling between FeSe electron and the STO phonon is responsible for the enhancement of T$_c$ over other FeSe-based superconductors[1,2]. However the recent observation of a similar superconducting gap in FeSe grown on the (110) surface of STO raises the question of whether a similar mechanism applies[3,4]. Here we report the ARPES study of the electronic structure of FeSe grown on STO(110). Similar to the results in FeSe/STO(001), clear replica bands are observed. We also present a comparative study of STO (001) and STO(110) bare surfaces, where photo doping generates metallic surface states. Similar replica bands separating from the main band by approximately the same energy are observed, indicating this coupling is a generic feature of the STO surfaces and interfaces. Our findings suggest that the large superconducting gaps observed in FeSe films grown on two different STO surface terminations are likely enhanced by a common coupling between FeSe electrons and STO phonons.**

The discovery of high temperature superconductivity in 1 UC FeSe/STO(001) continues to attract a great deal of interest[1]. In particular, recent mutual inductance measurement has shown an onset of diamagnetism at the same temperature (65K) when a gap emerges in ARPES, thus supporting the superconducting origin of the observed single particle energy gap[5]. Comparing with the 8 K superconducting transition temperature of bulk FeSe the T$_c$ of the 1 UC FeSe/STO(001) is almost an order of magnitude higher. This leads to a natural question regarding the cause of the T$_c$ enhancement.

Apparently, one contributing factor is electronic doping. Earlier works[6–8] of intercalated A$_x$Fe$_{2-y}$Se$_2$ (A=K, Tl, Cs, Rb, etc.) compounds have shown superconductivity with T$_c$ above 30K. Recent work on a bulk crystal, Li$_{1-x}$Fe$_x$OHFeSe, consisting of FeSe layers intercalated with Li$_{1-x}$Fe$_x$OH shows T$_c$~41 K[9,10]. On a different front, when the top surface of a non-superconducting 3 UC FeSe/STO(001) is coated with potassium (K), a superconducting gap is observed at 48K[11]. However, despite the similarity in the fermiology of these systems with that of the 1UC FeSe/STO(001), the T$_c$ of the later is still significantly higher. This suggests that electron doping alone is insufficient to account for the full enhancement of T$_c$ in 1 UC FeSe/STO(001).

The foremost telling clue concerning the origin of the extra T$_c$ enhancement in 1 UC FeSe/STO(001) comes from the ARPES observation of the "replica bands" at ~100 meV below the main band[2]. Such a replica band is explained in terms of a "shake off" phenomenon – in the photoemission process of ejecting an electron, part of the incoming photon energy can be used to excite a vibration quantum. Thus, the replica bands signify a strong coupling between the FeSe electron and this STO phonon, which in turn was conjectured to be the T$_c$ enhancement mechanism[2].

In two very recent papers scanning tunneling microscopy (STM) and ARPES reported similar superconducting gap and gap closing temperature of 1 UC FeSe grown on the (110) surface of STO[3,4]. Since the (110) surface is geometrically and chemically different from that of (001) surface, it raised doubts[4,12] over whether the same electron-phonon enhancement mechanism also applies.

We carried out ARPES studies of FeSe film on STO(110) to check this issue. In addition, we have carried out a comparative study of the electron-phonon coupling on bare STO(001) and STO(110) surfaces. Similar electron-phonon replica band has been found for an two-dimensional electron gas formed on the (001) surface[13,14]. Our result shows that the electron-phonon replicas at the FeSe/STO(110) interface and bare STO(110) surface are similar to its counterparts at the (001) surface, suggesting such electron-phonon coupling is a generic feature of STO surfaces and interfaces prepared under certain condition. As such, the recent observation in the FeSe/STO(110) system is consistent with the conjecture of $T_c$ enhancement by interface electron-phonon coupling.

**Results**

**Replica bands of 1UC FeSe grown on STO(110) and STO(001)** One important difference between the interfaces of FeSe/STO(110) and FeSe/STO(001) is the distortion of in-plane unit cell from tetragonal to orthorhombic symmetry. A comparative ARPES study of FeSe/STO(110) and FeSe/STO(001) has been reported recently by Zhang et al.[4]. However, the replica bands were not resolved in their experiment. By utilizing the capability of *in-situ* MBE growth and high resolution ARPES beamline, we have carefully studied the band strutures of the 1UC FeSe/STO(110). Fig. 1(a) shows a Fermi surface at the M point of the Brillouin zone measured at around 25 K. Fig. 1(b) shows the energy-momentum intensity map along a high symmetry cut through M. It clearly shows an electron band crossing the Fermi level. The superconducting gap opening at the Fermi momentum ($k_F$) is measured to be around 14 meV, similar to that observed in FeSe/STO(001)[2]. In order to better resolve band dispersion, we show the second energy derivatives of the ARPES intensity in Fig. 1(d), and mark the peaks of the energy distribution curves (EDCs) with green and red squares in Fig. 1(e). In Fig. 1(d) and Fig. 1(e) we see both the Fermi level crossing electron band and a hole band below it. Most interestingly, at ~100meV below the main bands we again see the features that can be identified as the replica bands. The replica of the hole band (red dash line) is clearly presented at Fig. 1(d), the totality of the data in Fig. 1(d) and 1(e) also makes the replica of the electron band (green dashed line) discernable. Sketches of the main bands and replica bands are shown by the solid and dashed curves, respectively in Fig.1(c). Thus replica bands with similar energy separation as in 1UC FeSe/STO(001) have been clearly identified in FeSe/STO (110).

**Surface electronic structure of STO(110) and STO(001)** To lend further support for the similarity of electron-phonon coupling on these surfaces and interfaces, we proceed with a comparative study the surface electronic structure of pure STO(110), in reference to (001) surface where the replica band has been recently being reported[13,14]. The two dimensional electron gas (2DEG) created by exposing the STO(001) surface to synchrotron radiation in ultrahigh vacuum, i.e., photo doping, has been extensively studied[15–17]. By tuning the dosage of UV exposure, which increases the electron density and by adding atomic oxygen which decreases the electron density, very precise control of the 2DEG density can be achieved[18,19]. Similar to STO (001), 2DEG has also been observed on the (110)[20] and (111)[19] surfaces of STO.

In particular, ARPES has observed the photo-doped surface conduction bands and their quantum well subbands due to spatial confinement.

The STO samples we used are wafers from the same source as the STO (001) substrate upon which we deposited the 1UC FeSe film in Ref. 2. Our results on the surface electronic structure are generally consistent with those reported by Wang et al.[20], as shown in Supplementary. We focus our discussion on lower photo doping density regime where replica is most visible. In Fig. 2(a) and (c) we compare the APRES data of STO(001) and STO(110) at electron density of $n_{2D} \approx$ $4.4 \times 10^{13}$ and $1.8 \times 10^{13}$ cm$^{-2}$, respectively. These electron densities are chosen so that the $k_F$ of the main bands are similar for the two surface terminations. In this doping regime we can only resolve the parabolic $d_{xy}$ main bands (the quantum well subbands cannot be resolved at this doping level). Importantly, a band with similar effective mass at higher binding energy (~100meV) can also be resolved for both surface terminations. This is the feature we identify as the replica bands. To enhance the contrast we plot the energy second derivative of the ARPES intensity in panels (b) and (d). The EDCs of STO(110) have also been shown as a waterfall plot at Fig. 2(e) to highlight the main band and the replica band. For both case we determine the energy separation between the main band and the replica band to be ~100meV. To more accurately estimate the energy separation and compare the relative intensity between the replica and main bands we show the EDCs at $k_F$ in Fig. 2(f). We have normalized the intensities so that the peaks associated with the main bands coincide for the two surface terminations. The deduced energy separation between the main band and the replica band are nearly the same.

**ARPES study of STO(110) as a function of photo doping** To further understand the electron-phonon coupling on STO surfaces, we explore its doping dependence. Figure 3 shows the band structure evolution of STO(110) with increasing photo doping. The carrier density increases from $n_{2D} \approx 1.8 \times 10^{13}$ cm$^{-2}$ to $6.0 \times 10^{13}$ cm$^{-2}$. The upper part of each panel shows the energy second derivative of lower part in the energy range of $-100$ meV $\leq E \leq 0$. As the carrier density increases the $d_{xy}$ band shifts toward higher binding energy, and its quantum well subband gradually appears, again resembling that of the STO(001) surface[14]. The subband is closely attached to the main band at first, then moves to lower binding energy and become clearly separated from the main band at higher carrier densities (the dashed yellow curve in Fig. 3(h)). The observation that the separation between the main band and the subband increases with doping is consistent with the quantum well origin of the subbands. In contrast the replica band is better resolved at low carrier densities (see, e.g., the red curve in Fig. 3(a)).

In Fig. 3(i) we show the doping dependence of the replica band intensity. Here the dashed red line marks the energy position of the replica band. Contrary to the quantum well subbands, the replica bands are better resolved at the lower doping level. We have estimated the intensity ratio between the replica and the main band. The result is plotted as a function of electron density in supplementary Fig. S2(a). The carrier density dependence described above are similar to those observed by Wang et al. on STO(001), consistent with a similar picture for electron-phonon coupling on STO surfaces. We also estimate the effective mass associated with the parabolic $d_{xy}$ bands at different carrier densities (Fig. S2(b)). If one attributes the change in the effective mass to the change in the electron-phonon coupling strength the result is consistent with the trend deduced from Supplementary Fig. S2(a).

We also observed a similar replica band with the same phonon mode on STO(111) surface at lower doping level (see Supplementary), again supporting the picture of common behavior of electron-phonon coupling in various STO surfaces and interfaces. This suggests the possibilities of high $T_c$ superconductivity in FeSe/STO(111) if good quality interface can be achieved.

**Discussion**

The origin of high $T_c$ in 1UC FeSe/STO is still a question under active current studies. From the experimental facts at least two contributing factors can be identified[21]: (1) the electron doping and (2) the substrate effect. The most direct evidence that electron doping raises $T_c$ are the studies of potassium doping on otherwise non-superconducting multilayer FeSe films (or low $T_c$ bulk FeSe). By coating the surface of bulk FeSe[22,23] and multi-layer FeSe films [11,24–27] with potassium $T_c$ can be raised to as high as 48K, while their Fermi surfaces have a similar area as that of 1UC FeSe/STO, hence approximately the same electron doping level. In addition, bulk materials where various donor layers are intercalated between the FeSe layers, e.g. $A_xFe_{2-y}Se_2$[28] and $Li_{1-x}Fe_xOHFeSe$[9] also have similar Fermi surface volume and similar range of $T_c$. Nonetheless, the $T_c$ of these systems are still appreciably lower than 1UC films on STO or BTO[29] (see Fig.4).

Thus it is natural to associate the extra enhancement of $T_c$ in the monolayer films to the substrate effect. In particular, the cross-interface coupling of the FeSe electrons and STO phonons has been proposed to enhance the $T_c$ of heavily electron doped FeSe systems[2,21]. On the surface it is unusual that electrons in the top layer of K-doped 3UC FeSe/STO cannot couple to the STO phonons, i.e., the electron-phonon coupling being very local.

To answer that question we recall that the doping in 1UC FeSe/STO is due to the charge transfer between STO and FeSe so that near the FeSe-STO interface there is an electric field. Such electric field will induce a layer of dipoles, so that near the interface with FeSe, STO is strongly polarized. In Ref. 2 and 21 it has been argued that the phonons causing the replica are associated with the vibration of these dipoles with the displacement perpendicular to the interface, and the modulating wave vector parallel to interface. Such in-plane modulating dipole potential does not affect layers further away from the interface due to the screening by the electrons in the bottom FeSe layer. The fact that the bottom layer alone can achieve screening suggests that Coulomb attraction localizes nearly all doped charges in the bottom FeSe layer. Thus while the top layer in K-doped 3UC FeSe/STO is equally electron doped (by potassium rather than STO) it does not experience the modulating dipole potential due to the screening of other FeSe layers beneath it.

The above situation has a close analogy with what we observed here for the doping dependence of the replica band on the surface of STO. As the photo doped carrier density increases, the modulating dipole potential caused by the polar phonon is screened. As a result the replica band intensity vanishes.

In conclusion, we have observed the replica band for FeSe/STO(110) as well as lightly doped STO(110) surface. The similar energy separation between the main bands and replica bands for these interfaces/surfaces as those in FeSe/STO(001) and STO(001) suggest the same electron-phonon coupling is involved in both. Our results suggest a similar phonon-enhancement mechanism might also be at work for the superconducting FeSe/STO(110).

## Methods

**Samples.** The pure substrates of STO were degassed at 450°C for 1 hour before in-situ transferred to the ARPES measurement. Single unit cell FeSe film was grown on 0.05wt% Nb doped STO(110) substrate. Before growing, the substrates were degassed at 450 °C for 45 min, and heated to 830°C for 30 min in the UHV chamber. The substrates were kept at 390°C during film growth. FeSe was obtained by co-evaporating Fe and Se with a flux ratio of 1:10. Post-annealing at 420°C for 3 hours was performed after growth. The growth rate was approximately 2UC/min.

**ARPES measurements.** All the measurements were performed in-situ at MBE+beamline ARPES systems in this work. The ARPES measurements of pure substrates were carried out at Beamline 10.0.1 of the Advanced Light Source (ALS) of Lawrence Berkeley National Laboratory. The total energy resolution was set ~25 meV with photon energy of 48 eV and the base pressure was better than $5 \times 10^{-11}$ Torr. The measurements on 1UC FeSe/STO(110) were performed at Beamline 5-4 of the Stanford Synchrotron Radiation Lightsource (SSRL) of SLAC National Accelerator Laboratory. The photon energy used was 24 eV with the total energy resolution ~8 meV and the base pressure was also better than $5 \times 10^{-11}$ Torr.

## Acknowledgements

The work at Stanford is supported by the US DOE, Office of Basic Energy Science, Division of Materials Science and Engineering, under award number DE-AC02-76SF00515. ALS and SSRL are supported by the Office of Basic Energy Sciences, U.S. DOE under contract No. DE-AC02-05CH11231 and DE-AC02- 76SF00515, respectively. CFZ's postdoctoral fellowship is supported by Knut and Alice Wallenberg Foundation in Sweden. DHL is supported by the U.S. Department of Energy, Office of Science, Basic Energy Sciences, Materials Sciences and Engineering Division, grant DE-AC02-05CH11231.

## Author contributions

C. Z., Y. H., R. M., H. H., D. L. and Z. X. S. proposed and designed the research. C. Z. and Z. L. carried out the ARPES measurements with help from Z. C., Y. X., J. H., S. T., T. J., W. L., H. Y. and S. R.. S. K. M., D. L., M. H. maintained the ARPES beamlines. C. Z. analyzed the data with help from Z. L. and E. M.. C. Z. wrote the manuscript with D. L., R. M., R. H., H. H. and Z. X. S.. R. M, H. H, D. L., Z. X. S. are responsible for project direction and infrastructure. All authors discussed the results and commented on the manuscript.

## Additional information

**Competing financial interests:** The authors declare no competing financial interests.

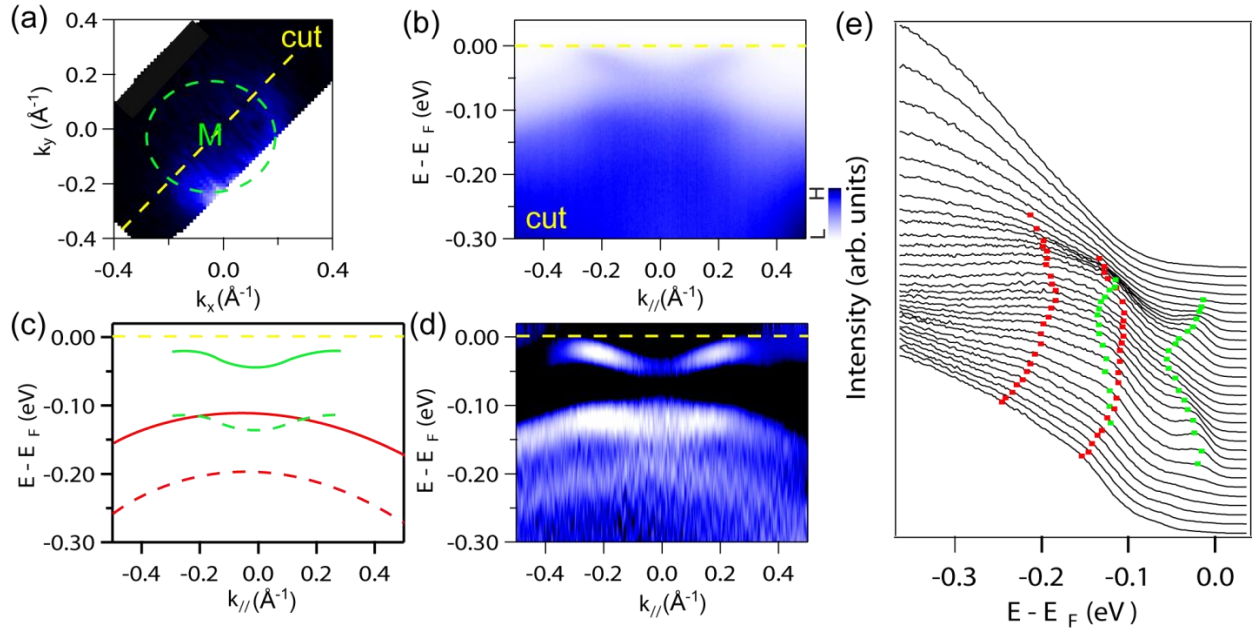

**Fig.1 The ARPES spectra of 1UC FeSe grown on STO(110). a** The M point Fermi surface map of FeSe/STO(110). **b** Energy-momentum intensity map of FeSe/STO(110) around a high symmetry cut through M. **c** Schematic representation of the electron band (green) and hole band (red) of FeSe/STO(110). The replica bands are shown as the dashed lines. **d** Second derivative image of **b**. Comparison with panel **c** identifies the features associated with the main bands and the replica bands. **e** EDCs near M shown as a waterfall plot with main and replica bands marked by corresponding color squares.

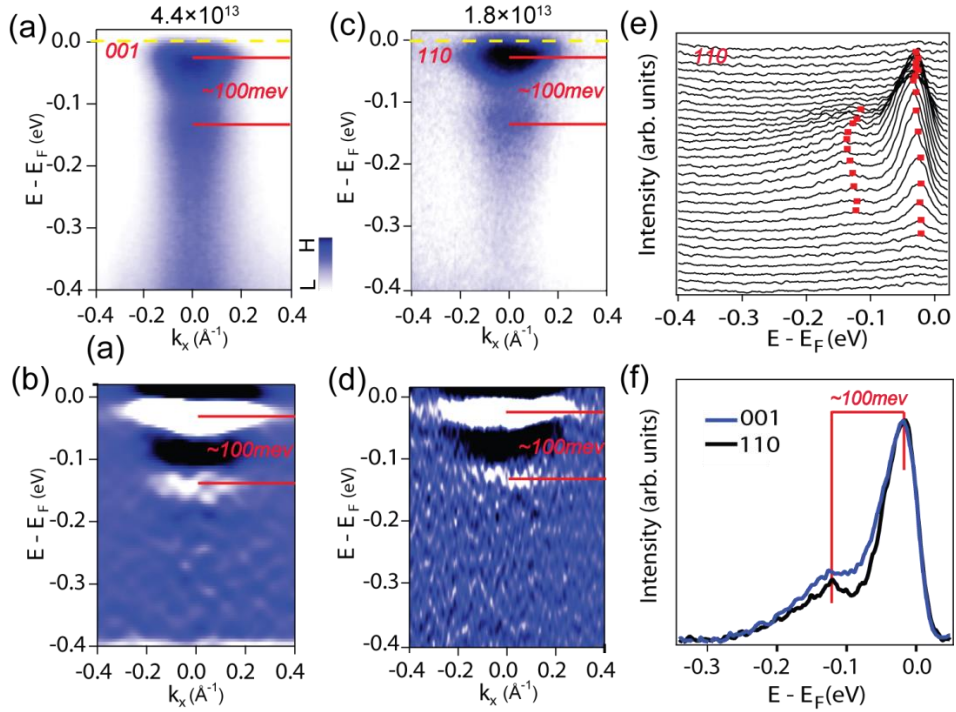

**Fig.2 The comparison at the ARPES spectra for the surface bands at STO(001) and STO(110). a, c** Energy-momentum intensity map of STO(001) and STO (110) with similar main bands and replica bands. **b, d** second energy derivative of **a** and **c**. **e** EDCs of STO(110) shown as a waterfall plot with main and replica bands marked by red squares. **f** Energy distribution curves of STO(001) and STO(110) measured at $k_F$.

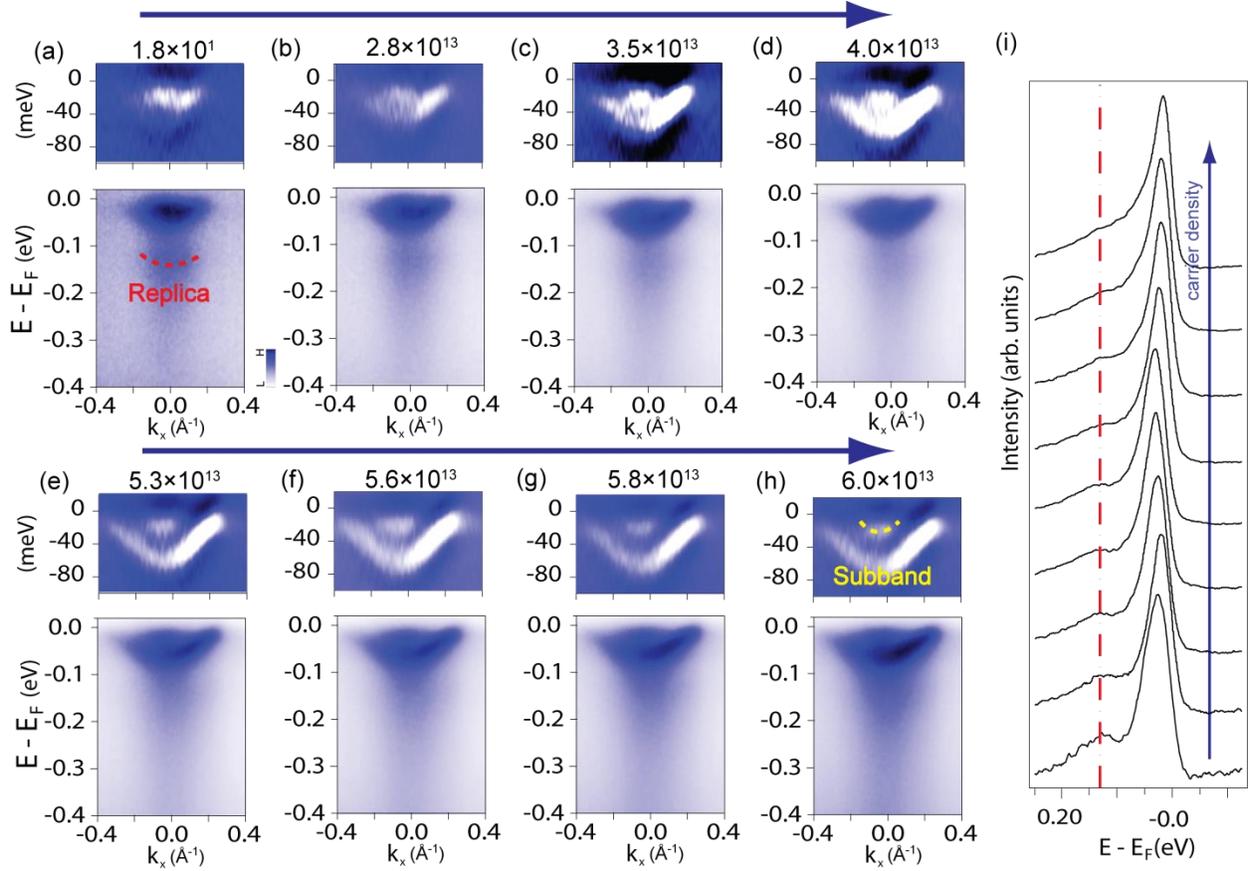

**Fig.3 The photodoping evolution of the ARPES spectra for the surface bands of STO(110). a-h** The evolution of the ARPES dispersion as the surface carrier density increases from $n_{2D} \approx 1.8 \times 10^{13}$ cm$^{-2}$ to $6.0 \times 10^{13}$ cm$^{-2}$. The red curve marks the replica band in panel(a) while the yellow curve marks quantum well subband in panel(h). The upper part of each panel shows the second energy derivative of the lower part in the -100 meV $\leq$ E $\leq$ 0 energy window. **i** energy distribution curves of the main band and replica band as the carrier increases from $n_{2D} \approx 1.8 \times 10^{13}$ cm$^{-2}$ to $6.0 \times 10^{13}$ cm$^{-2}$. The dashed red line marked the positon of replica band.

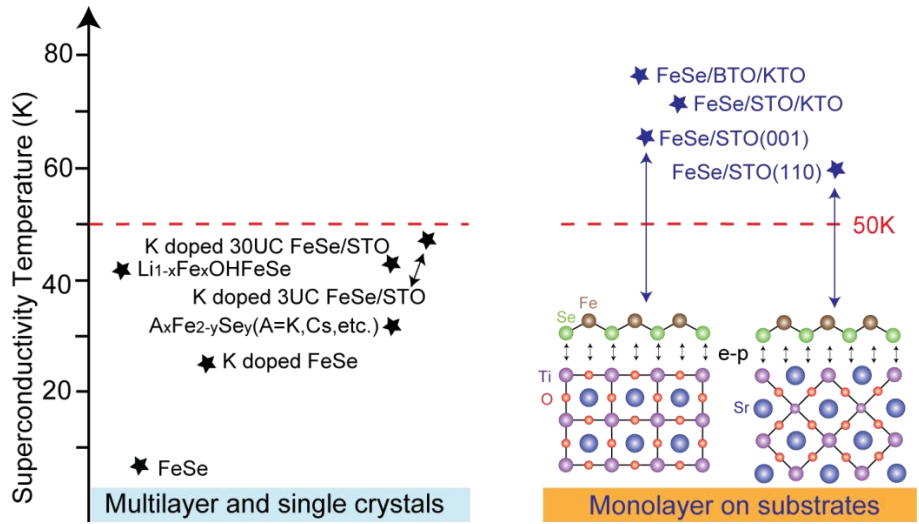

**Fig.4 Superconductivity temperature of FeSe related superconductors.** All the multilayer and single crystals of iron based superconductors show Tc lower than 50K. In contrast, monolayer FeSe on various $TiO_2$ terminated substrates always show Tc higher than 50K. We have observed electron-phonon coupling both at FeSe/STO(001) and FeSe/STO(110). The $T_c$ of FeSe/STO/KTO is from Ref. 30.